# Transverse Coupled Cavity VCSEL: Making 100 GHz Bandwidth Achievable

Moustafa Ahmed[1], Ahmed Bakry[1], Ahmed Alshahrie[1], and Hamed Dalir[2,*]

[1]Department of Physics, Faculty of Science, King Abdulaziz University, 80203 Jeddah 21589, Saudi Arabia

[2]Department of Electrical and Computer Engineering, George Washington University, 20052, Washington, D.C., USA

[*]Corresponding author: hdalir@gwu.edu

**Abstract**

Enhancing the modulation bandwidth (MBW) of semiconductor lasers has been the challenge of research and technology to meet the need of high-speed photonic applications. In this paper, we propose the design of vertical-cavity surface-emitting laser integrated with multiple transverse coupled cavities (MTCCs) as a promising device with ultra-high 3-dB bandwidth. The laser features high modulation performance because of the accumulated strong coupling of the slow-light feedback from the surrounding lateral TCCs into the VCSEL cavity. Photon-photon resonance (PPR) is predicted to occur at ultra-high frequencies exceeding 145 GHz due to the optical feedback from short TCCs, which achieves 3-dB MBW reaching 170 GHz. The study is based on the modeling of the VCSEL dynamics under multiple transverse slow-light feedback from the surrounding TCCs. We show that the integration of the VCSEL with four or six feedback cavities is advantageous over the TCC-VCSEL in achieving much higher MBW enhancement under weaker coupling of slow-light into the VCSEL cavity. We also characterize the noise properties of the promising MTCC-VCSEL in the regime of ultra-high bandwidth in terms of the Fourier spectrum of the relative intensity noise (RIN).

## 2. Introduction

Due to the unique features of VCSELs, such as high efficiency, low power consumption, high temperature stability, and direct manufacturing of dense arrays [1,2], light emitters using directly modulated VCSELs are attractive for cost-effective photonic applications. However, the 3-dB bandwidth of the VCSEL is limited to 30 GHz mainly due to the carrier-photon resonance (CPR), thermal effects, and parasitic resistance /capacitance [3,4]. Therefore, challenges have been put on the VCSEL to enhance its bandwidth to meet the requirements of speed beyond 100 Gb/s by the current applications of internet, supercomputers and data centers of increased network traffic [5]. In the last decade, different solutions were proposed to boost the transmission bitrate of the semiconductor lasers. External optical feedback was recognized as a technique to increase MBW of the cost-effective directly modulated semiconductor lasers [6,7]. Under strong feedback, the intensity modulation (IM) may excite PPR between the modulating field and the external-cavity oscillating modes [8-12]. This PPR is manifested in the modulation spectrum as a second resonant modulation peak beyond the intrinsic CPR peak, and hence enhancing the bandwidth [13,14].

It has been realized that application of external optical feedback can help to increase the MBW of VCSELs [8,9]. Dalir et al. [15-18] demonstrated that adding a single TCC to a primary VCSEL cavity can increase MBW. The stated design principle was to control the slow-light delay time in the TCC and the induced slow-light feedback, which implies the use of a PPR effect. Ahmed et al [19] showed that the strong PPR effect produces peaky structures in the modulation response, termed as "resonance modulation response". Such bandwidth enhancement effect is sensitive to the coupling strength variation. On the other hand, they reported that a shorter TCC is advantageous to boost the CPR effect and bandwidth, however, this requires strong coupling to achieve the same MBW enhancement as that by a long TCC [19].

Most recently, the first and last authors have taken part in demonstrating a novel VCSEL design of a hexagonal transverse-coupled-cavity adiabatically coupled through a central cavity [20], as shown in figure 1. Following this scheme, the authors showed a prototype demonstrating a 3-dB roll-off MBW of 45 GHz, which is five times greater than a conventional VCSEL fabricated on the same epiwafer structure [20]. The structure provided slow-light coupling into the VCSEL cavity more than the VCSEL with single TCC which then works to increase the IM response beyond the CPR frequency and boost the bandwidth [21]. These results simulate the authors to investigate further the modulation performance of MTTC-VCSEL and optimize the device structure toward further enhancement of MBW.

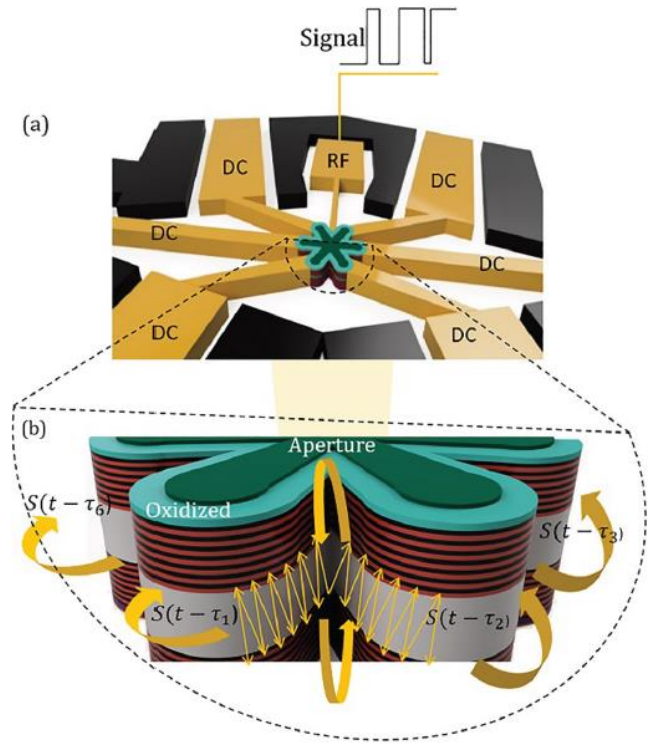

**Figure 1:** Schematic structure of our hexagonal transverse-coupled cavity vertical-cavity surface-emitting lasers (VCSEL) (a) Top view, and (b) Cross-sectional view [20].

In this paper, we present modeling of VCSELs coupled with different schemes of multi-lateral TCCs surrounding the VCSEL cavity toward boosting MBW further in the mm-waveband. We propose the VCSEL to be surrounded by the TCCs, which provides direct slow-light feedback from every TCC into the primary cavity. Therefore, even if the direct feedback from every TCC is weak or intermediate, it accumulates to couple more slow light into the VCSEL cavity and induce feedback strong enough to achieve more MBW enhancement. We introduce a theoretical model of the VCSEL dynamics under the induced multiple slow-light feedback, derive the corresponding threshold gain and modify the rate equations of the VCSEL. Basing on numerical integration of these time-delay rate equations, we simulate and compare the IM response of the VCSEL design using one, two, four and six TCCs. We elucidate the advantages of coupling the VCSEL with multiple TCCs over the single or double TCCs in not only achieving more MBW enhancement but also using much weaker optical coupling into the VCSEL cavity. We present results on ultra-high bandwidth exceeding 170 GHz using VCSEL integrated with four and six TCCs, which to the best of our knowledge, is the highest predicted value: thanks to the MTCC structure and the PPR effect. In addition, we characterize the noise properties of the 4TCCs-VCSEL and 6TCCs-VCSEL in the preferable regime of ultra-high bandwidth.

In the next section, we introduce the design and modeling of the of the MTCCs VCSEL. The procedures of numerical calculation are given in section 3. The results on the IM response with bandwidth enhancement and on noise properties are presented in section 4. Finally, we introduce the concluding remarks of the present work in section 5.

## 2. Model of Slow-Light Feedback in VCSEL due Multi-Surrounding TCCs

The proposed design and model of the MTTC-VCSEL are illustrated in the scheme of figure 2, which is a top view of the VCSEL surrounded in the lateral direction with multiple lossy cavities through oxide apertures. This structure of coupled waveguides introduces lateral optical confinement and a leaky traveling wave in the direction of each TCC. Unlike a conventional VCSEL design, light generated in the MTCC-based laser has additional lateral components with an angle close to 90° near the cutoff condition of light propagation [16]. That is, light travels perpendicularly and is slowed in the laterally coupled waveguides. Within each TCC, the slow light propagates for several round trips with group velocity of $v_g = c / n_g$, where $n_g = fn$ is the group index, $n$ is the average material refractive index and $f$ is the slow factor of light. The slow light is totally reflected back at the far end of the TCC and is coupled into the VCSEL cavity with a coupling ratio $\eta$. The back and forth propagating slow light of the $m^{th}$ TCC suffers a loss of $\exp(-2\alpha_{Cm}L_{Cm})$ and phase delay of $\exp(-2j\beta_{Cm}L_{Cm})$, where $\alpha_{Cm}$ and $\beta_{Cm}$ are the lateral optical loss and propagation constant with $\kappa$ being the material loss and $\lambda$ the emission wavelength. The period of the round trip between the VCSEL cavity and the far end of the $m^{th}$ TCC is $\tau_m = 2n_{gm}L_{Cm}/c$. In this case, the threshold gain level of the VCSEL cavity $G_{thc}$ is modified to the form.

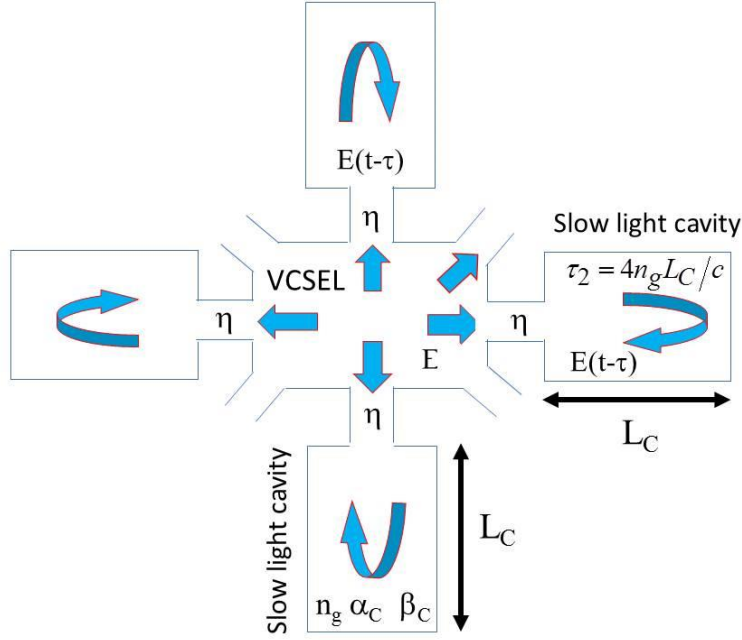

**Figure 2.** Schematic top-view of VCSEL with each of the multiple surrounding cavities providing direct slow-light feedback.

$$G_{th} = G_{th0} - \frac{v_g}{W} \ln \prod_{m=1}^{M} |U_m(t - \tau_m)| \tag{1}$$

which constitutes a generalized form for the gain derived in references [19,22]. The feedback function $U_m(t-\tau_m)$ is the time-delay function describing the slow-light feedback from the $m^{th}$ TCC.

$$U_m(t - \tau_m) = |U_m(t - \tau_m)| e^{j\varphi}$$
$$= 1 + \frac{\eta_m}{1 - \eta_m} \sum_p \sqrt{1 - \eta_m}^p e^{-2p\alpha_{Cm} L_{Cm}} e^{-j2p\beta_{Cm} L_{Cm}} \frac{E(t - p\tau_m)}{E(t)} \tag{2}$$

where the summation is over the multiple round trips in the TCC. In this case, $\theta(t - p\tau_m) - \theta(t)$ represents the deviation in the optical phase due to chirping in the $m$th cavity. The rate equations of the MTCC-VCSEL is given for the injected electron number $N(t)$, photon number $S(t)$ contained in the lasing mode and the optical phase $\theta(t) = \arg[E(t)]$ as:

$$\frac{dN}{dt} = \frac{\eta_i}{e} I(t) - av_g \frac{(N - N_T)}{1 + \varepsilon S} S - \frac{N}{\tau_e} + F_N(t) \tag{3}$$

$$\frac{dS}{dt} = \left[\Gamma a v_g \frac{(N-N_T)}{1+\varepsilon S} - \frac{1}{\tau_p} + \frac{v_g}{W}\sum_{m=1}^{M}\ln|U(t-\tau_m)|\right]S + \Gamma R_{sp} + F_S(t) \quad (4)$$

$$\frac{d\theta}{dt} = \frac{\alpha}{2}\left(\Gamma a v_g(N-N_{th}) - \frac{v_g}{W}\sum_{m=1}^{M}\varphi_m\right) + F_\theta(t) \quad (5)$$

where $a$ is the differential gain of the active region whose volume is $V$, $N_T$ the electron numbers at transparency, and $\varepsilon$ the gain suppression coefficient. $\Gamma$ is the confinement factor, $\tau_p = 1/G_{thD}$ is the photon lifetime, $\eta_i$ is the injection efficiency, $\tau_s$ is the electron lifetime due to the spontaneous emission, $R_{sp}$ is the spontaneous emission rate, and $N_{th}$ is the electron number at threshold. In equation (3), the injection current is assumed to have sinusoidal current modulation with bias component $I_b$, modulation component $I_m$, and modulation frequency $f_m$. The Langevin noise sources in Eqs. (1) – (3) are given in respective by [23]

$$f_S(t) = \sqrt{\frac{2R_{sp}\Gamma S(t)}{\Delta t}}.x_S \quad (6)$$

$$f_N(t) = \sqrt{\frac{2N(t)}{\tau_s\Delta t}}.x_n - \sqrt{\frac{2R_{sp}S(t)}{\Delta t}}.x_N \quad (7)$$

$$f_\theta = \sqrt{\frac{R_{sp}\Gamma}{2S(t)\Delta t}}.x_\theta \quad (8)$$

where $x_s$, $x_N$ and $x_\theta$ are noise sources having normal distributions with zero mean and variance of unity. The frequency content of intensity fluctuations is measured in terms of RIN, which is calculated from the fluctuations $\delta S(t) = S(t) - S_b$ in $S(t)$, where $S_b$ is the bias value of $S(t)$. Over a finite time $T$, RIN is given as [24]

$$RIN = \frac{1}{S_b^2}\left\{\frac{1}{T}\left|\int_0^T \delta S(t)e^{-j2\pi f\tau}d\tau\right|^2\right\} \quad (9)$$

where $f$ is the Fourier frequency.

## 3. Numerical Calculations

In the present calculations and for simplicity, we assumed that the TCCs are identical, each TCC has length $L_C$, group index $n_g$, propagation constant $\beta_C$, optical loss $\alpha_C$, and hence round trip $\tau$. Also, the lateral slow light is coupled to the primary VCSEL cavity with equal coupling ratio $\eta$. Therefore, the threshold gain in equation (1) is reduced to

$$G_{th} = G_{th0} - M \frac{v_g}{W} \ln|U(t-\tau)| \tag{10}$$

with the feedback function $U(t-\tau)$ id then given by

$$\begin{aligned} U(t-\tau) &= |U(t-\tau_m)|e^{j\varphi} \\ &= 1 + \frac{\eta}{1-\eta} \sum_p \sqrt{1-\eta}^{\,p}\, e^{-2p\alpha_C L_C}\, e^{-j2p\beta_C L_C} \sqrt{\frac{S(t-p\tau)}{S(t)}}\, e^{j\theta(t-p\tau)-j\theta(t)} \end{aligned} \tag{11}$$

We integrate rate equation (3) – (5) by means of the fourth order Runge-Kutta algorithm using a time step as short as $\Delta t = 0.2$ ps. The integration is first done for the solitary VCSEL ($\eta = 0$) between $t = 0$ and $\tau$, and the obtained values of $S(t)$ and $\theta(t)$ are then used for further integration of the time delayed version of the rate equations. The data sampled for characterization of laser dynamics and noise are collected after the laser operation is stabilized. We apply the numerical values of the VCSEL parameters given in table 1 [25]. The slow factor and material absorption loss are set to be $f = 40$ and the bias current is $I_b = 2$ mA. For the simulation of the IM response of the MTCC-VCSEL, we applied the fast Fourier transform (FFT) to the modulated laser signal as,

$$\text{IM – repsonse} = a_1(f_m)/a_1(f_m \to 0) \tag{12}$$

where $a_1(f_m)$ is the fundamental peak of the FFT spectra of the laser intensity at the modulation frequency $f_m$. In this calculation, we dropped the noise sources in rate equations (3) – (5).

Table 1. Definition and numerical values of the VCSEL parameters [25]

| Parameter | Value |
|---|---|
| Refractive index of active region $n$ | 3.3 |
| Material loss of active region $\alpha_m$ | 1000 m$^{-1}$ |
| Slow factor $f$ | 40 |
| Volume $V$ | 1.76x10$^{-19}$ m$^3$ |
| Width of the VCSEL cavity $W$ | 4 μm |
| Differential gain $a$ | 3.64x10$^{-12}$ m$^3$s$^{-1}$ |
| Confinement factor $\Gamma$ | 0.0382 |
| Electron number at transparency $N_T$ | 3.17x10$^5$ |
| Gain suppression coefficient $\varepsilon$ | 2.25x10$^{-5}$ s$^{-1}$ |
| Photon lifetime $\tau_{ph}$ | 2 ps |
| Spontaneous emission rate $R_{sp}{'}$ | 6.6x10$^{28}$ m$^{-3}$s$^{-1}$ |
| Injection efficiency $\eta_i$ | 0.6 |
| Electron lifetime $\tau_s$ | 1.5 ns |
| Linewidth enhancement factor $\alpha$ | 2 |

## 4. Results and Discussions

A. Modulation response of VCSEL with single TCC

Examples of the IM response spectra with improved MBW $f_{3dB}$ of a VCSEL integrated with single TCC are plotted in figure 3. These spectra correspond to TCC length of $L_C$ = 5 μm and coupling ratio of $\eta$ = 0.75, 0.8 and 0.9. The figure indicates an increase of MBW from that of the conventional VCSEL (C-VCSEL) $f_{3dB0}$ = 21.5 GHz to $f_{3dB}$ = 27.5, 36 and 40 GHz in the TCC VCSEL when $\eta$ = 0.75, 0.8 and 0.9, respectively. This range of $\eta$ corresponds to very strong transverse feedback. The increase of MBW can be attributed to strong anti-phase coupling between the transverse coupled radiation and the vertically lasing radiation in the VCSEL cavity [16].

The predicted values of MBW $f_{3dB}$ as a function of the coupling ratio $\eta$ are plotted in figure 4 for TCC lengths of $L_C$ = 5 and 6 μm. The figure indicates the increase of $f_{3dB}$ in the regime of very strong feedback with coupling ratios of $\eta$ > 0.71 when $L_C$ = 5 μm and $\eta$ > 0.78 when $L_C$ = 6 μm. The values of $f_{3dB}$ are smaller when $L_C$ = 6 μm than those when $L_C$ = 5 μm since the strength of the optical feedback decreases with the increase of the length of the feedback waveguide.

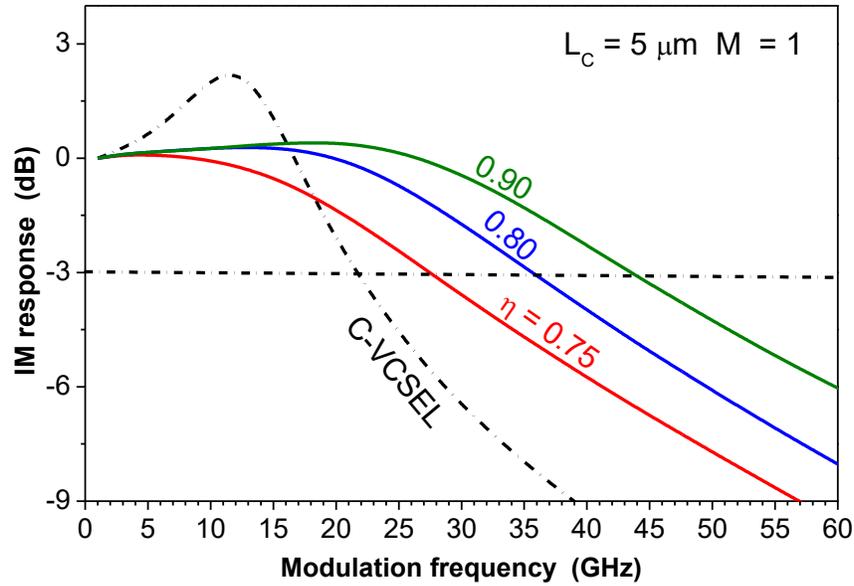

**Figure 3.** IM response with bandwidth improvement of the design of TCC VCSEL with $L_C = 5$ μm when $\eta = 0.75$, 0.8 and 0.9. The IM response of the C-VCSEL is plotted for comparison with the dashed-dotted line.

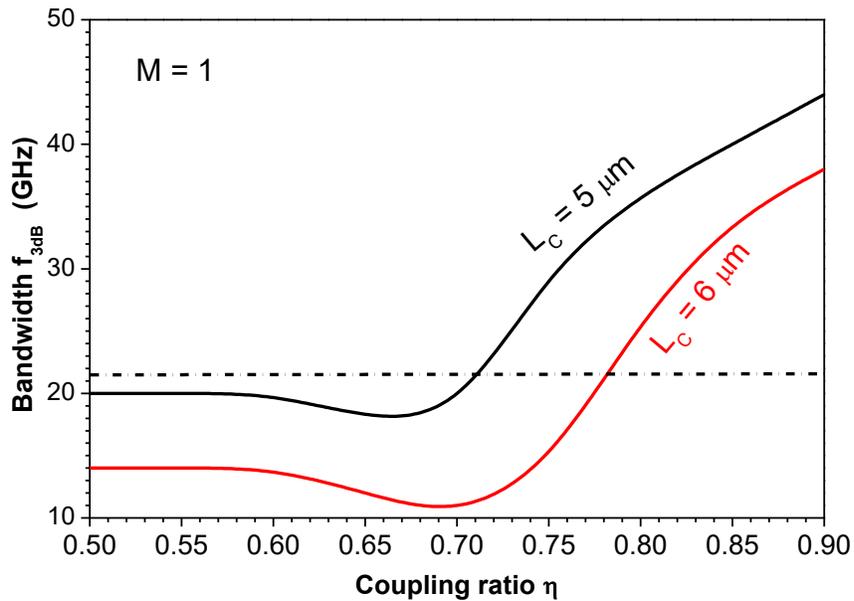

**Figure 4.** Variation of the bandwidth $f_{3dB}$ of the TTC VCSEL with the coupling ratio η when $L_C = 5$ and 6 μm.

B. Modulation response of MTCC-VCSEL

In this subsection, we show the trend how adding more lateral TCCs not only lowers the regime of coupling that corresponds to MBW improvement but also enhances this bandwidth to much higher and interesting values. Figures 5(a) – (c) plot examples of the IM responses with bandwidth enhancement when the number of lateral TCCs increases to M = 2, 4 and 6, respectively. All cases correspond to TCC length of $L_C$ = 5 μm. Figure 5(a) of M = 2 shows that the MBW values of $f_{3dB}$ = 40 and 45 GHz are obtained in this case when $\eta$ = 0.4 and 0.5, respectively, which are almost one half the values in figure 3 that results in the same bandwidth in the TCC-VCSEL. The bandwidth increases further to 60 and 61 GHz when the coupling ratio increases to $\eta$ = 0.7 and 0.9, respectively. It is worth noting that a PPR effect is induced with the increase of $\eta$; a PPR peak is seen around modulation frequency $f_{PP}$ = 180 GHz when $\eta$ = 0.9. This PPR is a result of modulation at frequencies close to the beating frequencies of external cavity oscillating modes [8-12]. When the VCSEL is surrounded by 4 TCCs (M = 4), figure 5(b) shows that the low values of the coupling ratio of $\eta$ = 0.2 and 0.3 accumulate so more slow-light feedback in the VCSEL cavity that MBW is increased to $f_{3dB}$ = 48 and 64, respectively. The PPR effect is remarkable when $\eta$ = 0.5 which is seen as resonant modulation with PPR peak of 5.8 dB around a very high frequency of $f_{PP}$ = 176 GHz. In this case, MBW is enhanced to $f_{3dB}$ = 80 GHz. The further increase of the light coupling into the VCSEL cavity to $\eta$ = 0.6 compensates to the loss in the drop of the IM response under the -3dB level and results in an ultra-high MBW of $f_{3dB}$ = 170 GHz associated with a high PPR peak around $f_{PP}$ = 145 GHz. These values, to the best of our knowledge, are the highest predicted values to the semiconductor laser bandwidth. Figure 5(c) indicates more improvement of the modulation performance with higher values of MBW at lower values of the coupling ratio $\eta$ when the number of TCCs increases to M = 6. At the low coupling of $\eta$ = 0.15, the bandwidth of the 6TTC-VCSEL is $f_{3dB}$ = 56 GHz. The PPR effect is initiated at lower values of the coupling ratio; when $\eta$ = 0.3 a PPR peak is seen around frequency $f_{PP}$ = 175 GHz. The associated bandwidth is $f_{3dB}$ = 76 GHz. When $\eta$ increases to 0.45, PPR is too enhanced to reveal resonant modulation of 6.2dB around frequency $f_{PP}$ = 165 GHz and the associated bandwidth of $f_{3dB}$ = 85 GHz. In this case of 6TCC-VCSEL, the ultra-high bandwidth of $f_{3dB}$ = 165 GHz is obtained at lower coupling of $\eta$ = 0.5, which is associated also with higher PPR peak around $f_{PP}$ = 147 GHz.

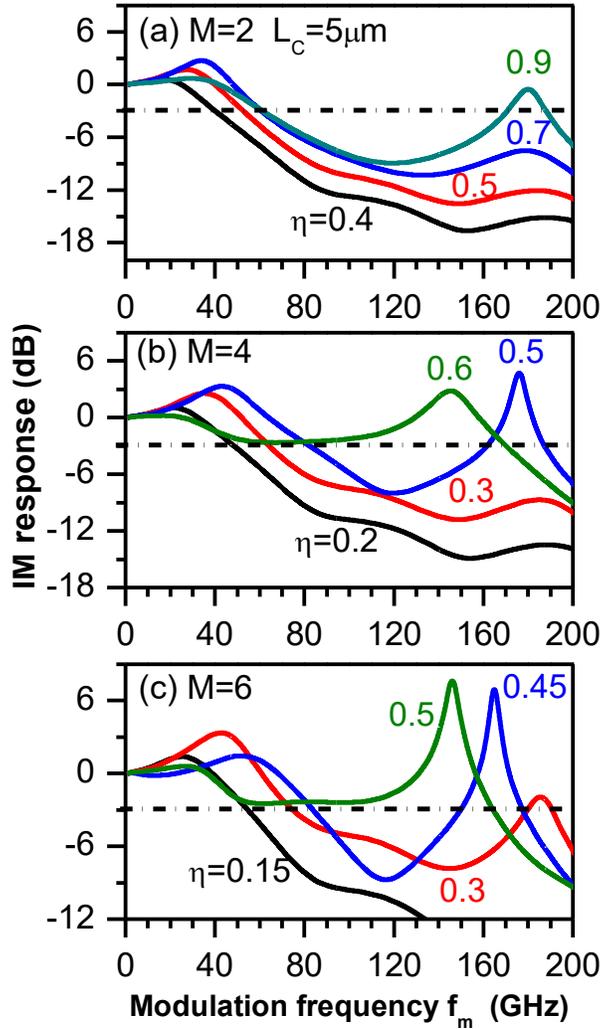

**Figure 5.** IM response of the design of (a) 2TCC-VCSE, (b) 4TCC-VCSE and (c) 6TCC-VCSE with $L_C$ = 5μm at different values of $\eta$ that results in MBW enhancement.

In figure 6(a), we plot variation of MBW $f_{3dB}$ with the coupling ratio $\eta$ for the three cases of MTCC-VCSEL (M = 2, 4 and 6) when $L_C$ = 5 μm. The figure shows that the bandwidth improvement when M = 2 is initiated when $\eta$ > 0.35 and $f_{3dB}$ reaches 62 GHz when $\eta$ > 0.75. In case of the 4TCC VCSEL, the bandwidth is much more enhanced when $\eta$ > 0.15. When $\eta$ > 0.4 the PPR effect is remarkable and the bandwidth reaches $f_{3dB}$ = 84 GHz when $\eta$ = 0.64. In the range of 0.65 ≤ $\eta$ ≤ 0.77, the strong light feedback works to recover the gap between the CPR and PPR peaks above the -3 dB level and the bandwidth is much more enhanced to values between $f_{3dB}$ = 88 and $f_{3dB}$ = 178 GHz. The bandwidth improvement is achieved at weaker feedback coupling of

$\eta > 0.10$ by the 6TTC-VCSEL. When $\eta > 0.30$, the boosted PPR effect is associated with enhanced bandwidth to values reaching $f_{3dB} = 84$ GHz. The further increase of $\eta$ between 0.45 and 0.65 raises the IM response above the -3dB level, which is smaller than the range achieved by the 4TTC-VCSEL. In this case, the predicted bandwidth values are in the ultra-high frequency range of $f_{3dB} = 88 \sim 180$ GHz. The corresponding variations of $f_{3dB}$ with $\eta$ when $L_C = 6$ μm are plotted in figure 6(b). The figure shows that the behaviors of $f_{3dB}$ with $\eta$ for the three cases of M = 2, 4 and 6 are similar in general to those when $L_C = 5$ μm. However, the increase of the bandwidth of the MTCC-VCSEL above that of the C-VCSEL occurs at smaller values of the coupling ratio $\eta$, and the predicted bandwidth is lower. The highest bandwidth ranges between $f_{3dB} = 80$ and 155 GHz.

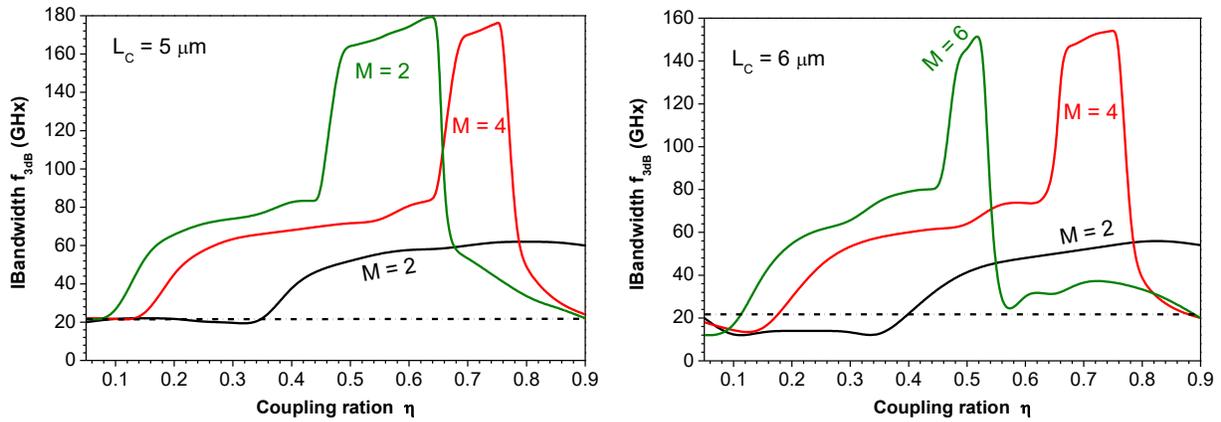

**Figure 6.** Variation of the bandwidth $f_{3dB}$ of the TTC VCSEL with the coupling ratio η when (a) $L_C = 5$μm and (b) $L_C = 6$μm for 2TCC-VCSEL, 4TCC-VCSEL and 6TCC-VCSEL.

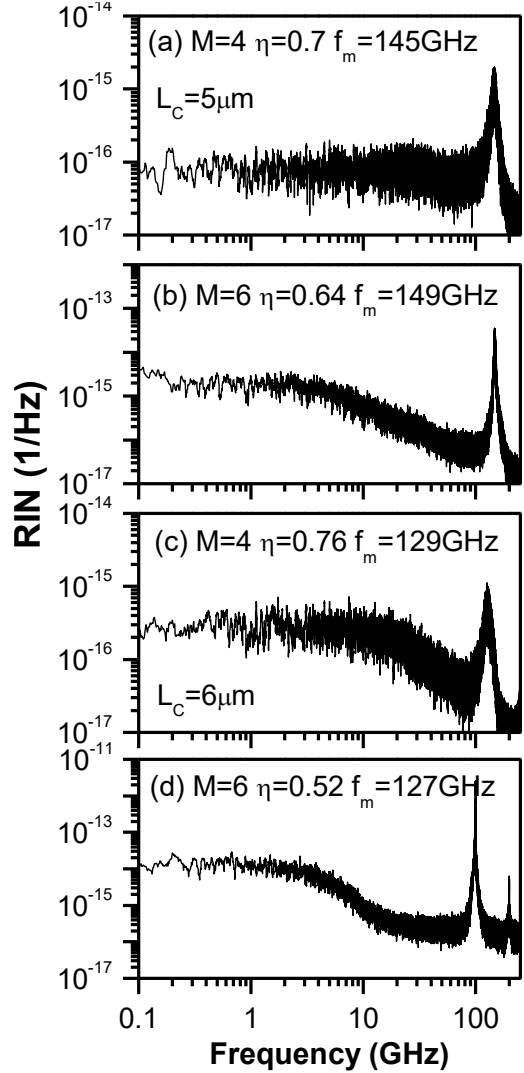

**Figure 7.** RIN spectra of MTCC-VCSEL with $L_C = 5$ μm when (a) M = 4, η = 0.7, $f_m$ = 145 GH, (b) M = 6, η = 0.64, $f_m$ = 149 GHz, and of MTCC VCSEL with $L_C = 6$ μm when (c) M = 4, η = 0.76, $f_m$ = 129 GHz and (d) M = 6, η = 0.52, $f_m$ = 127 GHz.

C. Noise properties of MTTC VCSEL

In this subsection, we investigate the noise properties of the MTCC-VCSEL under current modulation. We focus on the regime of ultra-high bandwidth, $f_{3dB} > 100$ GHz, shown in figure 6 for cases of VCSEL integrated with four TCCs (M=4) and six TCCs (M = 6). The noise is evaluated in terms of the frequency spectrum of RIN when the MTCC-VCSEL is modulated with a modulation frequency $f_m$ equal to one of the PPR frequencies in the regimes of ultra-high bandwidth. As explored in figure 6, these regimes are $0.65 \leq \eta \leq 0.77$ for 4TCC-VCSEL with $L_C$ = 5 µm, respectively, while they are $0.45 \leq \eta \leq 0.65$ for 4TCC-VCSEL with $L_C$ = 6 µm. Figure 7(a) – (d) plot four examples of the simulated RIN spectra of MTCC-VCSEL with $L_C$ = 5 µm when (M = 4, η = 0.7, $f_m$ = 145 GH) and when (M = 6, η = 0.64, $f_m$ = 149 GHz), and of MTCC-VCSEL with $L_C$ = 6 µm when (M = 4, η = 0.76, $f_m$ = 129 GHz) and when (M = 6, η = 0.52, $f_m$ = 127 GHz), respectively. The figures show that the RIN spectra exhibit pronounced peaks around the modulation frequency $f_m$ due to the high degree of periodicity. At frequencies lower than the regime of the resonance peak, the RIN level increases in general with the decrease of frequency and then exhibits flat (white) noise in the regime of low frequencies $f_m < 1$ GHz. The figures indicate also that the level of the low-frequency RIN (LF-RIN) of the 4TTC-VCSEL is almost one-order of magnitude lower than that of the 6TCC-VCSEL. This difference may indicate that the modulated signal of the 6TCC-VCSEL at this ultra-high frequency is little more irregular than that of the 4TCC-VCSEL.

In figures 8(a) and (b), we compare variations of the LF-RIN level of the RIN spectra of the 4TCC-VCSEL and 6TCC-VCSEL, respectively, when the TCC length is $L_C$ = 5 and 6 µm. The figure indicates an increase of LF-RIN with the increase of the coupling ratio $\eta$ in general. The noise levels when $L_C$ = 5 µm are comparable to those of the non-modulated TCC VCSEL as investigated by Ibrahim et al. [23]. Moreover, the figure shows that the noise levels when $L_C$ = 6 µm are higher than those when $L_C$ = 5 µm. Finally, the noise levels in the 6TCC-VCSEL are almost one-order of magnitude higher than those in the 4TCC-VCSEL over the same range of slow-light feedback. However, the investigated ranges of LF-RIN of the MTTC-VCSEL are still much lower than the level $> 10^{-8}$ 1/Hz that characterizes the unstable dynamics of the VCSEL [23].

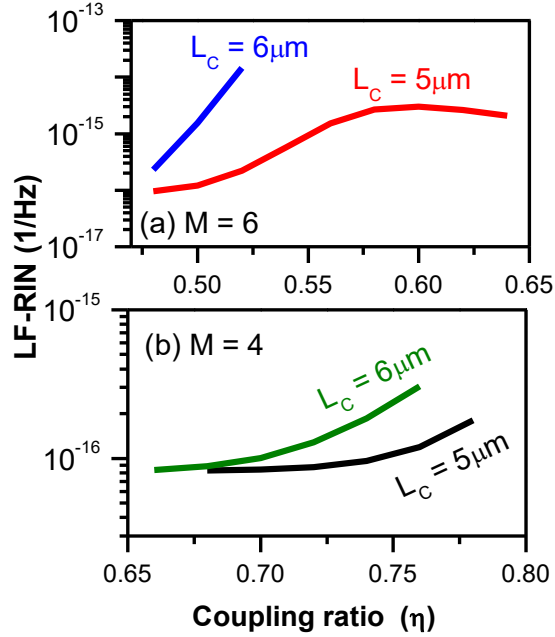

**Figure 8.** Variation of the LF-RIN level of (a) the 4TCC-VCSEL an(b) 6TCC-VCSELwhen the TCC length is $L_C$ = 5 and 6 μm.

Before concluding, we will briefly discuss this paper results in the wider context of advancements in nanophotonic emitters and laser [26]. The interplay between EM field, the gain feedback, and laser performance is indeed forming an intricate system comprised of the optical mode (or few-mode) cavities [27-29] and laser physics [30-37], both at the nanoscale and (sub)diffraction limited optical modes. Our work demonstrating enhanced laser modulating speed performance can also be seen as an extension of the ongoing discussion in the field of miniaturized laser devices. Here the debate around whether the Purcell factor, which captures the light-matter-interaction strength of such as of a laser cavity being proportional to the cold-cavities' quality factor (Q) divided by the cavities' mode volume, has (or has not) an influence on both the PPR and the gain relaxation frequencies, i.e. speed of the laser [30], is an open debate to date. The Purcell factor is especially high in light emitters and lasers featuring a sub diffraction limited optical mode, primarily due to the nonlinear scaling of volume and introduced loss due to the cavities' inability to provide feedback. This impacts the PPR in such small-volume cavities at (or below) the diffraction-limit of light [30-33], because the laser design with enhanced $F_p$ are also capable of increasing the temporal relaxations oscillations of the laser cavity, thus expanding the 'speed' of the laser under direct modulation. For instance, and example of this is the transverse-coupled cavity laser design providing coherent feedback from a plurality of cavities, thus enhancing the light emission from a central lasing cavity [20]. Looking ahead, future research should also investigate the effects of PPR in cavities with high longitudinal modes, for example in fiber optic-based laser systems for Brillouin amplification [38]. As it stands, the coherent feedback design of opportunely engineering a multiple TCC-based laser, as shown here, offers a new degree of freedom in laser and VCSEL design explorations. Given the predicted performance, these emerging VCSELs are poised to show a significant impact next-generation 5- and 6G network systems, data-centers, and high-end sensors systems.

## 5. Conclusions

In conclusion, promising designs, and modeling of VCSEL surrounded in the lateral direction by multiple TCCs was presented for the enhancement of MBW. We showed that the coupled slow light from each of the surrounding TCCs are accumulated in such a way to provide strong feedback in the VCSEL cavity stronger than the cases of VCSEL coupled with single TCC or two TCCs. This strong feedback works to enhance both the PPR and modulation bandwidth to values to values around 150 GHz in VCSELs coupled with four and six transverse cavities. The RIN spectrum is pronounced around the modulation frequency while the low-frequency part is flat with levels comparable to the non-modulated TCC VCSEL.

## Acknowledgment

This project was funded by the Deanship of Scientific Research (DSR) at King Abdulaziz University, Jeddah, under grant no. (**RG-18-130-41**). The authors, therefore, acknowledge with thanks DSR technical and financial support.

**References**

[1] K. Iga, "Vertical-cavity surface-emitting laser: Its conception and evolution," *Jpn. J. Appl. Phys.* **47**(1R), 1–10 (2008).

[2] F. Koyama, "Recent advances of VCSEL photonics," *J. Lightw. Technol.* **24**(12), 4502–4513 ,2006.

[3] L. A. Coldren, S. W. Corzine, and M. L. Mashanovitch, "Diode Lasers and Photonic Integrated Circuits" vol. **218**, Hoboken, John Wiley & Son, p. 260, (2012).

[4] C. Wilmsen, H. Temkin, and L. A. Coldren, "Vertical-cavity Surface- Emitting Lasers: Design, Fabrication, Characterization, and Applications" vol. **24**, United Kingdom, Cambridge University Press (2001).

[5] S. Kajiya, K. Ksukamoto, and S. Komaki, "Proposal of fiber-opticradio highway networks using CDMA method," *IEICE Trans. Electron.*,vol. **E79-C** (1), pp. 496–500 (1996).

[6] U. Feiste, "Optimization of modulation bandwidth of DBR lasers with detuned Bragg reflectors, "IEEE *J. Quantum Electron*. **34**, pp. 2371–2379 (1998).

[7] R. Mindaugas, A. Glitzky, U. Bandelow, M. Wolfrum. U. Troppenz, J. Kreissl, and W. Rehbein., "Improving the Modulation Bandwidth in Semiconductor Lasers by Passive Feedback" *IEEE J. Sel. Top. Quantum Electron.* **13**, pp. 136–142 (2007).

[8] A. Paraskevopoulos, H. J. Hensel, W. D. Molzow, et al., in OFC Conference and the National Fiber Optic Engineers Conference, paper PDP22 (2006).

[9] C. Chen and K. D. Choquette, "Analog and digital functionalities of coupled cavity surface emitting lasers," *J. Lightwave Technol*. **28**, pp. 1003–1010 (2010).

[10] P. Westbergh, J. S. Gustavsson, B. Kögel, Å. Haglund, and A. Larsson, "Impact of photon lifetime on high-speed VCSEL performance*," IEEE J. Sel. Top. Quant*. **17**(6) pp. 1603–1613 (2011).

[11] H. Dalir, A. Matsutani, M. Ahmed, A. Bakry, and F. Koyama, "High frequency modulation of transverse-coupled- cavity VCSELs for radio over fiber applications," *IEEE Photon. Technol. Lett.* **26**(3), 281-283 (2014).

[12] P. Bardella, W. Chow and I. Montrosset, "Design and Analysis of Enhanced Modulation Response in Integrated Coupled Cavities DBR Lasers Using Photon-Photon Resonance," *Photon*. 3, 1-14 (2016).

[13] M. S. Alghamdi, H. Dalir, A. Bakry, R. T. Chen, and M. Ahmed, "Regimes of Bandwidth Enhancement in Coupled-Cavity Semiconductor-Laser Using Photon-Photon Resonance," *Jpn. J. Appl. Phys*. **58**(11), 112003(6pp) (2019).

[14] H. Dalir, A. Matsutani, M. Ahmed, A. Bakry, and F. Koyama, "High frequency modulation of transverse-coupled- cavity VCSELs for radio over fiber applications," *IEEE Photon. Technol. Lett*. **26**(3) 281-283 (2014).

[15] H. Dalir, and F. Koyama, "Modulation bandwidth enhancement of VCSELs with lateral optical feedback of slow light", in *Proceedings of IEEE International Semiconductor Laser Conference* (IEEE, 2010), 83-84.

[16] H. Dalir and F. Koyama, "Bandwidth enhancement of single-mode VCSEL with lateral optical feedback of slow light," *IEICE Electron. Express* **8**(13), 1075-1081 (2011).

[17] H. Dalir and F. Koyama, "29 GHz directly modulated 980 nm vertical-cavity surface emitting lasers with bow-tie shape transverse coupled cavity," *Appl. Phys. Lett*. **103**(9) 091109 (2013).

[18] H. Dalir and F. Koyama, "High speed operation of bow-tie-shaped oxide aperture VCSELs with photon-photon resonance*,*" *Appl. Phys. Express* **7**, 022102 (2014).

[19] M. F. Ahmed, A. Bakry, R. Altuwirqi, M. S. Alghamdi, and F. Koyama, "Enhancing the modulation bandwidth of VCSELs to the millimeter-waveband using strong transverse slow-light feedback". *Opt Express*. **23**(7), 15365–15371 (2015).

[20] E. Heidari, H. Dalir, M. Ahmed, V. J. Sorger and R. Chen, "Hexagonal transverse-coupled-cavity VCSEL redefining the high-speed lasers", Nanophotonics **9**(16), pp. 4743–4748 (2020).

[21] M. Ahmed, H. Dalir and R. T. Chen, "Optical Devices with Transverse-Coupled Cavity," USA Patent and Trademark Office, Patent US010658815B1 (2020).

[22] R. Lang and K. Kobayashi, "External optical feedback effects on semiconductor injection laser properties," *IEEE J. Quantum Electron*. **16**(3), 347–355 (1980).

[23] H. R. Ibrahim, M. S. Alghamdi, A. Bakry, M. Ahmed and F. Koyama, "Modelling and characterisation of the noise characteristics of the vertical cavity surface-emitting lasers subject to slow light feedback," *Pramana Journal of Physics*, **93**, 73(7pp) (2019).

[24] M. Ahmed, M. Yamada, and M. Saito, "Numerical modeling of intensity and phase noise in semiconductor lasers", *IEEE J. Quantum Electron*. **37**, 1600 (2001).

[25] L. A. Coldren, S. W. Corzine, and M. L. Mashanovitch, *Diode lasers and photonic integrated circuits*, **2nd ed**. New York, NY: Wiley, p. 260 (2012).

[26] A. Fratalocchi, C. M Dodson, R. Zia, P. Genevet, E. Verhagen, H. Altug, et al. "Nano-optics gets practical" *Nature Nanotechnology* **10**, 11-15 (2015).

[27] V. J. Sorger, R. F. Oulton, J. Yao, G. Bartal, and X. Zhang "Fabry-Perot Plasmonic Nanocavity," *Nano Letters* **9**, 3489-3493 (2009).

[28] B. Min, E. Ostby, V. J. Sorger, E. Ulin-Avila, L. Yang, X. Zhang, and K. Valhalla "High-Q surface plasmon-polariton whispering-gallery microcavity," *Nature* **457**, 455-458 (2009).

[29] K. Liu, N. Li, D. K. Sadana, and V. J. Sorger "Integrated nano-cavity plasmon light-sources for on-chip optical interconnects," *ACS Photonics* **3**, 233-242 (2016).

[30] R. F. Oulton, V. J. Sorger, T. Zentgraf, R.-M. Ma, C. Gladden, L. Dai, G. Bartal, and X. Zhang, "Plasmon Lasers at Deep Subwavelength Scale," *Nature* **461**, 629-631 (2009).

[31] R.-M. Ma, R. F. Oulton, V. J. Sorger, and X. Zhang "Room-temperature sub-diffraction-limited plasmon laser by total internal reflection," *Nature Mat.* **10**, 110-113 (2010).

[32] R.-M. Ma, R. F. Oulton, V. J. Sorger, and X. Zhang "Plasmon lasers: coherent light source at molecular scales," *Laser & Photonics Reviews* **7**, 1-21 (2012).

[33] R.-M. Ma, X. Yin, R. F. Oulton, V. J. Sorger, and X. Zhang "Multiplexed and Electrically Modulated Plasmon Laser Circuit," *Nano Letters* **12**, 5396-5402 (2012).

[34] N. Li, K. Liu, V. J. Sorger, and D. K. Sadana, "Monolithic III-V on Silicon Nanolaser structure for optical Interconnects," *Scientific Reports* **5**, 14067 (2015).

[35] V. J. Sorger, "2D Material Nanowatt Threshold Lasing," *Nature Materials* **7**, e200 (2015).

[36] K. Liu, and V. J. Sorger "Electrically-driven Carbon nanotube-based plasmonic laser on Silicon," *Optical Materials Express* **5**, 1910-1919 (2015).

[37] K. Liu, S. Sun, A. Majumdar, and V. J. Sorger "Fundamental Scaling Laws in Nanophotonics" *Scientific Reports* **6**, 37419 (2016).

[38] F. S. Gokhan, H. Goktas, and V. J. Sorger "Analytical approach of Brillouin amplification over threshold" *Applied Optics* **57**(4), 607-611 (2018).